\documentclass{CHEP2004}

\usepackage{graphicx}

\newcommand{\etal}{{\it et al. }}

\begin{document}
%
\voffset -37 mm
\title{Multi-Terabyte EIDE Disk Arrays running Linux RAID5}

\author{D. A. Sanders, L. M. Cremaldi, V. Eschenburg, R. Godang, M. D. Joy and 
D. J. Summers, \\ University of Mississippi, University, MS 38677, USA\\
 \\
D. L. Petravick, FNAL, Batavia, IL 60510, USA}

\maketitle

\begin{abstract}
High-energy physics experiments are currently recording large amounts of data 
and in a few years will be recording prodigious quantities of data.  New methods 
must be developed to handle this data and make analysis at universities possible. 
Grid Computing is one method; however, the data must be cached at the various 
Grid nodes.  We examine some storage techniques that exploit recent developments 
in commodity hardware.  Disk arrays using RAID level 5 (RAID-5) include both parity 
and striping.  The striping improves access speed.  The parity protects data in the 
event of a single disk failure, but not in the case of multiple disk failures.

We report on tests of dual-processor Linux Software RAID-5 arrays and Hardware 
RAID-5 arrays using a 12-disk 3ware controller, in conjunction with 250 and 300 GB 
disks, for use in offline high-energy physics data analysis.  The price of IDE disks is 
now less than \$1/GB.  These RAID-5 disk arrays can be scaled to sizes affordable 
to small institutions and used when fast random access at low cost is important.
\end{abstract}

\section{INTRODUCTION}
We have tested redundant arrays of integrated drive electronics (IDE) disk drives, 
using the Linux operating system, for use in particle physics Monte Carlo simulations 
and data analysis. Parts costs of total systems using commodity IDE disks are now 
at the \$2000 per terabyte level. A revolution is in the making. Our tests include reports 
on Software and Hardware redundant arrays of inexpensive disks -- Level 5 
(RAID-5)  systems running under Linux . RAID-5 protects data in case of a 
catastrophic single disk failure by providing parity bits. Journaling file systems 
are used to allow rapid recovery (minutes rather than days) from system crashes 
and power failures. 

Our data analysis strategy is to encapsulate data and CPU processing power 
together. Data is stored on many PCs. Analysis of a particular part of a 
data set takes place locally on, or close to, the PC where the data resides. 
The network backbone is only used to put results together. 
If the I/O overhead is moderate and analysis tasks need more than one 
local CPU to plow through data, then each of these disk arrays could be 
used as a local file server to a few computers sharing a local network switch. 
These commodity network switches would be combined with a 
single high end, fast backplane switch allowing the connection of a thousand 
PCs. To this end, we have also successfully measured the file transfer speed of  
Network File System (NFS) software over a local Gigabit network.

RAID \cite{RAID} stands for Redundant Array of Inexpensive Disks.  Many 
industry offerings meet all of the qualifications except the inexpensive part, 
severely limiting the size of an array for a given budget. This is now changing. 
The different RAID levels can be defined as follow:
\begin{itemize}
\item RAID-0: ``Striped.'' Disks are combined into one physical device where 
reads and writes of data are done in parallel. Access speed is fast but there 
is no redundancy.
\item RAID-1: ``Mirrored.'' Fully redundant, but the size is limited to 
the smallest disk.
\item RAID-4: ``Parity.'' For $N$ disks, 1 disk is used as a parity bit and 
the remaining $N-1$ disks are combined. Protects against a single disk 
failure but access speed is slow since you have to update the parity disk for 
each write. Some, but not all, files may be recoverable if two disks fail.
\item RAID-5: ``Striped-Parity.'' As with RAID-4, the effective size 
is that of $N-1$ disks. However, since the parity information is also 
distributed evenly among the $N$ drives the bottleneck of having to update 
the parity disk for each write is avoided. Protects against a single disk 
failure and the access speed is fast.
\end{itemize}

Hardware and Software RAID-5, using enhanced integrated drive electronics 
(EIDE) disks under Linux software, is now available \cite{RAID-5}. Redundant 
disk arrays do provide protection in the most likely single disk failure case, that 
in which a single disk simply stops working. This removes a major obstacle to 
building large arrays of EIDE disks. However, RAID-5 does not totally protect 
against other types of disk failures. RAID-5 will offer limited protection in the 
case where a single disk stops working but causes the whole EIDE bus to fail 
(or the whole EIDE controller card to fail), but only temporarily stops them 
from functioning. This would temporarily disable the whole RAID-5 array. 
If replacing the bad disk solves the problem, i.e.~the failure did not 
permanently damage data on other disks, then the RAID-5 array would recover 
normally. 

\section{TEST SETUP}
To get a large RAID array one needs to use large capacity disk drives. There 
have been some problems with using large disks, primarily the maximum 
addressable size. We have discussed these problems in an earlier 
papers \cite{CHEP98, IEEE, CHEP03}. Using arrays of disk drives, such as those 
shown in Table \ref{Disks}, one can create Multi-Terabyte RAID arrays.
\begin{table*}[t!]
\begin{center}
\caption{Comparison of Large EIDE Disks for a RAID-5 Array}
\tabcolsep=2.0mm
\begin{tabular}{lcccccr}\hline
Disk Model&Size (GB)&RPM&Cost/GB&GB/platter&Cache Buffer&Warranty\\ 
 \hline
Maxtor Maxline II \cite{maxline}& 300& 5400& \$0.75& 75&2 MB& 3 year\\
Western Digital WD2500JB \cite{WDC250}& 250& 7200& \$0.61& 80&8 MB& 3 year\\
&&&&&&\\
Maxtor MaXLine\,Plus\,II \cite{maxline}& 250& 7200& \$0.69& 80&8 MB& 3 year\\
Maxtor MaXLine\,Plus\,III  SATA\cite{maxline3}& 300& 7200& \$0.77& 100&8 MB& 3 year\\
Seagate Barracuda SATA \cite{seagate}& 200& 7200& \$0.63& 100&8 MB& 5 year\\
Hitachi DeskStar 7K400 \cite{IBM400} & 400 & 7200& \$1.10& 80&8 MB& 3 year\\
\hline
\end{tabular}
\label{Disks}
\end{center}
\end{table*}

To test both the Software and Hardware RAID-5 arrays we started with a base 
system with two different modifications. The base system consisted of the following: 
120 GB Western Digital system disk, a MSI K7D Master MPX motherboard, a dual 
2 GHz AMD Athlon for the CPUs, with 1024 MB DDR memory, a Gigabit Ethernet 
card. We also needed to use a second Power  Supply (15A at 12V) to supply 
enough power at startup for all the disks. The startup power needed was about 
450 Watts.  we also needed 24 inch EIDE Cables and , due to space and air-flow 
considerations, we use the round EIDE cables rather than the ribbon cables. 
(Base cost: \$1400)

For Software RAID-5 we used either eight 250 GB Western Digital disks \cite{WDC250} 
or eight 300 GB Maxtor disks \cite{maxline} and 2 Promise Ultra133 PCI (Peripheral Component Interconnect) EIDE controller cards \cite{promise}. We originally planed to try 
using 3 Promise cards but we found the there were too many PCI Interrupt request conflicts. 
This limited us to 8 disks, which was convenient since we would have otherwise run into the 
2 Terabyte "disk" size limit of Linux kernel 2.4 at this point. (Cost: Base plus  \$2300 for a 
total of \$3700 or about \$1850/TB.)

For Hardware RAID-5 we used twelve 250 GB Western Digital disks and the 3ware 
12-disk RAID controller 7506-12 \cite{3ware}. This controller also had a 2 Terabyte "disk" 
size limit. To overcome this limit we had to split the 12 disks into 2 Hardware RAID-5 arrays 
of six disks, each forming a 1.2 TB RAID-5 array. We then combined them, using Software 
RAID-0 into a 2.4 TB array. We then upgraded to the Linux 2.6 kernel, which does support 
arrays larger than 2 Terabytes ($2^{32}$ 512 byte blocks) when used with the newer 3ware 
Escalade 9500S-12 controller but not our 7506-12. (Cost: Base plus \$3800 for a total 
of \$5200 or about \$1890/TB.)

\section{RESULTS}
After confirming the fact that the CPU swapping algorithms do allow for efficient use of 
dual-CPU computers (kswap was typically $10-15\%$ for CPU intensive jobs), we tested 
the array write speeds with a simple program that wrote $3.28\times 10^9$ Bytes of plain 
text (``All work and no play make Jack a dull boy'' ). We had the following results:

\subsection{Software RAID-5}
For the Software RAID-5 we only used 8 of the 250 GB disks, thus we were under the 
2 Terabyte "disk" size limit of Linux kernel 2.4. Using the test program described above 
we had a base (with only that job running) write speed of 29 MB/s. For two concurrent 
writes (2 instances of the the same job writing to 2 different files) we had a speed of 
24 MB/s, an overhead of $17 \%$. For a read test we copied the file to the system disk 
for a rate of 37 MB/s. When simply copying the file back to the RAID-5 array we had a 
write speed of 33 MB/s.  The CPU overhead of journaling, Software RAID, and writing 
the file was about $10-15\% $, but for a single instance write this was running on the 
other CPU. Therefore, for fast CPUs, the overhead of Software RAID-5 is negligible.

 \subsection{Hardware RAID-5}
Because of the 2 Terabyte "disk" size limit of Linux kernel 2.4, we first tried using nine 
250 GB disks (out of the 12 possible disks) and Hardware RAID-5, forming a 2 TB array. 
The base write speed, as described above, was 41 MB/s. We then used Linux kernel 2.6 
so we could have a larger RAID array, however, we discovered that the Hardware RAID 
controller also had a 2 TB "disk" size limit. Therefore we created 2 Hardware RAID-5 
arrays of six disks, each forming a 1.2 TB RAID-5 array. We then combined them using 
Software RAID-0 into a 2.4 TB array. We now had a base write speed of 29 MB/s. 
For two concurrent writes we had a speed of 25 MB/s, an overhead of $14 \%$. We then 
turned off the RAID-0 and performed some additional speed tests. Again, when simply 
copying the file back to the RAID-5 array we had a write speed of 33 MB/s.  When we 
copied the file from one Hardware RAID-5 to another the speed was 37 MB/s. The CPU 
overhead of journaling and writing the file was about $1-5\% $, making the Hardware 
RAID-5 array more efficient but, if one is using the array as only a disk server then even 
a single CPU would suffice.

\subsection{Gigabit Network}
To test the practical speed of a local Gigabit network we connected the Software RAID-5 
array and another Linux computer together using an inexpensive 8-port commodity 
Gigabit switch \cite{Dlinkswitch}. We then first mounted the array via NFS. When mounted 
using synchronous NFS we had a write speed of 13 MB/s and when mounted using 
asynchronous NFS we had a write speed of 18 MB/s.  When simply copying the file to the 
RAID-5 array we had a write speed of 23 MB/s. These speeds should be compared with 
the base internal write speed of 29 MB/s and 33 MB/s for copying. The combined network 
and NFS overhead is $38\%$ for asynchronous writing and $30\%$ for copying. When we 
tried simple FTP we had a speed of 22 MB/s, a network overhead of $33\%$. 

\subsection{Motherboard Performance}
One should note that in previous tests \cite{CHEP03} we did see much higher writing speeds. 
For those tests we used a different motherboard \cite{asus} (we were only testing 
single-CPU arrays) that was noted for efficiently bridging the PCI bus.

\section{FUTURE RECOMMENDATIONS}
Over the last 3 years we have put together 5 RAID-5 arrays of various 
sizes \cite{IEEE, CHEP03}. The arrays have been used at SLAC on {\sc BaBar} and at 
CERN for CMS Monte Carlo \cite{CMS-Note}. This totaled 40 EIDE disks. Over $25\%$ 
failed within 3 years, fortunately within the warranty period. Some of this rate may be 
attributed to power failures, or perhaps a bad batch, but it still seems to be too high a 
rate. Given this failure rate and other considerations we would consider making the 
following recommendations. If you plan to build-it-yourself you should use hot-swappable 
Serial Advanced Technology Attachment (SATA) drives with at least a 3-year warranty. 
The connectors for SATA drives also take up far less space and they can safely operate 
over longer cable lengths than standard EIDE drives. This is important if you try to either 
use a tall ``Tower'' case or even a double-wide ``Tower'' case. The double-wide ``Tower'' 
case has the advantage of better airflow for cooling. We have used Supermicro 
CSE-733T-450B cases to  build boxes with hot swappable SATA disks in non-RAID5 systems.

If you want to use 12 disks you will need a Hardware RAID controller similar to the 3ware 
Escalade 9500S-12 (or 9500S-12MI) SATA card. You will want a 2.0 GHz AMD Athlon 
(or better) CPU. To connect to your local area network (LAN) of processing computers 
you will want at least Gigabit Ethernet, either a card or on the motherboard. You might 
also consider using a Fiber Channel Arbitrated Loop (FC-AL). FC-AL nominally runs at 
100 or 200 MB/s and can be daisy-chained between computers or connected to Fiber 
Channel switches. An FC-AL PCI card typically costs \$500 and comes with two ports for 
connection to a simple loop or to a fabric switch.  A fabric switch  typically costs \$15000 
for 16 ports and allows more simultaneous connections for increased speed. To exceed 
the 2 TB ``disk'' size limit you will also need to use Linux Kernel 2.6 and a journaling file 
system such as ext3 \cite{ext3} or ReiserFS \cite{Reiser}. ReiserFS created the journal 
faster than ext3 but it is not supported by RedHat Enterprise Level 3 version of Linux.

There are also various commercial RAID systems that rely on a hardware RAID controller. 
Examples of these are shown in Table \ref{Commercial}. They are typically 3U or larger 
rack mounted systems.  In the past, commercial systems have not been off-the-shelf 
commodity items.  This is changing and while they are anywhere from 1.5 to over 
eighteen times as expensive, even allowing for cost of assembly, having an off-the-shelf 
unit is quite attractive.  The Apple FC-AL Xserve RAID interface runs with both Macintosh 
and Linux computers.
\begin{table}[ht]
\begin{center}
\tabcolsep=1.5mm
\caption[]{Some Commodity Hardware RAID Arrays\cite{Apple}.}
\label{Commercial}
\begin{tabular}[t]{lrrr}\hline
System &Capacity&Size&Price/GB\\
\hline
Apple Xserve RAID & 3.5 TB &3U& \$3.14 \\
Dell EMC CX200 & 2.1 TB &3U& \$9.05 \\
HP StorageWorks 1000 & 2.1 TB &3U& \$11.39 \\
IBM FASt200 3542-1R & 2.1 TB&3U& \$24.71 \\
Sun StorEdge 6120 & 2.04 TB&2$\times $3U& \$36.57 \\
\hline
\end{tabular}
\end{center}
\end{table}

\section{CONCLUSION}
We have tested redundant arrays of IDE disk drives for use in 
offline high energy physics data analysis and Monte Carlo simulations. 
Parts costs of total systems using commodity IDE disks are now at 
the \$2000 per terabyte level. We have tested Software RAID-5 systems 
running under Linux 2.4 using Promise Ultra 133 disk controllers and Hardware 
RAID-5 systems running under Linux 2.4 and 2.6 using a 3ware Hardware RAID 
controller. We found about $5\%$ overhead for journaling files systems such as 
ext3 and ReiserFS, but given the extra protection and increased recovery speed 
we still recommend them. We also found that Software RAID-5 has about $10\%$ 
more overhead than Hardware RAID but the use of dual-CPU systems or using the 
RAID-5 array as a dedicated file server make this ``cost'' negligible for any modern 
CPU. RAID-5 provides parity bits to protect data in case of a single catastrophic 
disk failure. Tape backup is not required for data that can be recreated with modest 
effort. Journaling file systems permit rapid recovery from system crashes and power 
failures. 

Current high energy physics experiments, for example {\sc BaBar} at SLAC, feature 
relatively low data acquisition rates, only 3 MB/s, less than a third of 
the rates taken at Fermilab fixed target experiments a decade ago 
\cite{farm}. The Large Hadron Collider experiments CMS and ATLAS, 
with data acquisition rates starting at 100 MB/s, will be more challenging 
and require physical architectures that minimize helter skelter data movement 
if they are to fulfill their promise. In many cases, architectures designed 
to solve particular processing problems are far more cost effective than 
general solutions \cite{farm,E769}. 

Grid Computing \cite{grid} will entail the movement of large amounts of 
data between various sites. RAID-5 arrays will be needed as disk caches 
both during the transfer and when it reaches its final destination to ameliorate 
Grid-lock. Another example that can apply to Grid Computing is the Fermilab 
Mass Storage System, Enstore \cite{enstore}, where RAID arrays are used as 
a disk cache for a Tape Silo. Enstore uses RAID arrays to stage tapes to disk 
allowing faster analysis of large data sets.

\section{ACKNOWLEDGEMENTS}

Many thanks to S. Bracker, J. Izen, L. Lueking, R. Mount, M. Purohit, 
W. Toki, and T. Wildish for their help and suggestions.  
This work was supported in part by the U.S. Department of Energy under 
Grant Nos. DE-FG05-91ER40622 and DE-AC02-76CH03000.



\begin{thebibliography}{99} 

\bibitem{RAID}
D.~A.~Patterson, G.~Gibson and R.~H.~Katz,
Sigmod Record {\bf 17}, 109 (1988).

\bibitem{RAID-5}
M. de Icaza, I. Molnar, and G. Oxman,  ``The linux raid-1,4,5 code,'' 
in  \textit{3rd Annu. Linux Expo'97}, (April 1997).

\bibitem{maxline}Maxtor.
 (2003) Maxline ATA.  \hfill \break
http://www.maxtor.com/en/documentation/ \hfill \break
data{\_}sheets/maxline{\_}data{\_}sheet.pdf [2004].

\bibitem{WDC250}Western Digital Corp.
 (2003) Specifications for the  \hfill \break
 WD Caviar WD2500JB. http://www.wdc.com/en/products/ \hfill \break
current/drives.asp?Model=WD2500JB [2004]. 
 
 \bibitem{maxline3}Maxtor.
 (2004) MaXLine III.  \hfill \break
http://www.maxtor.com/{\_}files/maxtor/en{\_}us/documentation/ 
data{\_}sheets/maxline{\_}iii{\_}data{\_}sheet.pdf  [2004].

\bibitem{seagate}Seagate. (2004) Barracuda 7200.7 SATA.  \hfill \break 
http://www.seagate.com/cda/products/discsales/personal/ \hfill \break 
family/0,1085,599,00.html  [2004].

\bibitem{IBM400}Hitachi Global Storage Technologies. 
 (2004)  \hfill \break 
 Deskstar 7K400.  http://www.hitachigst.com/hdd/support/ \hfill \break
 7k400/7k400.htm  [2004].

\bibitem{CHEP98}
D.~Sanders, C.~Riley, L.~Cremaldi, D.~Summers and \hfill \break
D.~Petravick, 
 in \textit{Proc. Int. Conf. Computing in High- 
Energy Physics (CHEP 98)}, Chicago, IL, (Aug. 31 - Sep 4 1998) 
 [arXiv:hep-ex/9912067].

\bibitem{IEEE}
D.~A.~Sanders, L.~M.~Cremaldi, V.~Eschenburg, \hfill \break
C.~N.~Lawrence, C.~Riley, D.~J.~Summers and \hfill \break
D.~L.~Petravick,
IEEE Trans.\ Nucl.\ Sci.\  {\bf 49}, 1834 (2002)
[arXiv:hep-ex/0112003].

\bibitem{CHEP03}
D.~A.~Sanders {\it et al.},
eConf {\bf C0303241}, TUDT004 (2003)
[arXiv:physics/0306037].

\bibitem{promise}Promise Technologies, inc. 
 (2001) Ultra133 TX2 -- \hfill \break
 Ultra ATA/133 Controller for 66 MHz PCI Motherboards. 
http://www.promise.com/marketing/\hfill \break
datasheet/file/U133{\_}TX2{\_}DS.pdf  [2003] and \hfill \break
http://www.promise.com/marketing/datasheet/\hfill \break
file/Ultra133tx2DS{\_}v2.pdf [2003].\hfil \break
Each ATA/PCI Promise card controls four disks.

\bibitem{3ware}
3ware.  (2003) Escalade  7500 Series ATA RAID Controller. 
http://www.3ware.com/products/\hfill \break 
pdf/Escalade7500SeriesDS1-7.qk.pdf [2003]; \hfill \break 
3ware.  (2003) Escalade 7500-12 ATA RAID \hfill \break
Controller. http://www.3ware.com/ \hfill \break 
products/pdf/12-PortDS1-7.qk.pdf [2003].

\bibitem{Dlinkswitch}D-Link Systems. 
(2001) DGS - 1008T.  \hfill \break
http://www.dlink.com/products/switches/\hfill \break
dgs1008t/dgs1008t.pdf [2003].

\bibitem{asus}ASUS. 
 (2002) ASUS A7M266. \hfill \break
http://www.asus.com/mb/socketa/a7m266/overview.htm\hfill \break
 [2003]. This used the AMD 761 North Bridge chip set.

\bibitem{CMS-Note}
V.~Lefebure and T.~Wildish,
``The spring 2002 {DAQ} {TDR} production,'' CERN-CMS-NOTE-'\hfill \break
2002-034. See http://cmsdoc.cern.ch/\hfill \break
documents/02/note02{\_}034.pdf [2003].

\bibitem{ext3}A. Morton.
 (2002) ext3 for 2.4. \hfill \break
http://www.zip.com.au/$\sim $akpm/linux/ext3/ [2003].

\bibitem{Reiser}H. Reiser.
 (2001) Three reasons why ReiserFS is great for you. 
 http://www.reiserfs.org/ [2003].

\bibitem{Apple} Apple Computers.
 (2004) Xserve RAID. \hfill \break
 http://images.apple.com/server/pdfs/\hfill \break
 L301297A{\_}XserveRAID{\_}TO.pdf. [2004]\hfill \break
 (Based on suggested retail Prices on December 10, 2003)

\bibitem{farm}
For  example, a decade ago the Fermilab E791 collaboration recorded and
reconstructed 50 TB of raw data in order to generate charm physics
results. For details of the saga, in which more data was written to tape than
in all previous HEP experiments combined, see:  \hfill \break
S.~Amato, J.~R.~de Mello Neto, J.~de Miranda, C.~James, D.~J.~Summers 
and S.~B.~Bracker,
Nucl.\ Instrum.\ Meth.\ A {\bf 324}, 535 (1993) [arXiv:hep-ex/0001003]; \hfill \break
S.~Bracker and S.~Hansen, 
[arXiv:hep-ex/0210034];\hfill \break
S. Hansen, D. Graupman, S. Bracker and S. Wickert,
IEEE Trans.\ Nucl.\ Sci.\  {\bf 34}, 1003 (1987);\hfill \break
S.~Bracker, K.~Gounder, K.~Hendrix and D.~Summers,
IEEE Trans.\ Nucl.\ Sci.\  {\bf 43}, 2457 (1996)
[arXiv:hep-ex/9511009];\hfill \break
E. M. Aitala {\it et al.} [E791 Collaboration], 
Phys Lett. B {\bf 440}, 435 (1998) [arXiv:hep-ex/9809026];\hfill \break
E. M. Aitala {\it et al.} [E791 Collaboration], 
Eur. Phys. J. direct  C {\bf 1}, 4 (1998) [arXiv:hep-ex/9809029].

\bibitem{E769}
For a description of Fermilab's first UNIX farm:  \hfill \break
C.~Stoughton and D.~J.~Summers,
Comput.\ Phys.\  {\bf 6}, 371 (1992),
[arXiv:hep-ex/0007002]; \hfil \break
C.~Gay and S.~Bracker,
IEEE Trans.\ Nucl.\ Sci.\  {\bf 34}, 870 (1987);\hfill \break
G. A. Alves {\it et al.} [E769 Collaboration], 
Phys. Rev. Lett. {\bf 69}, 3147 (1992).

\bibitem{grid}
P.~Avery,
Phil.\ Trans.\ Roy.\ Soc.\ Lond.\  {\bf 360}, 1191 (2002); 
L. Lueking \etal, Lect.\ Notes\ Comput.\ Sci.\  {\bf 2242}, 177 (2001).

\bibitem{enstore}Fermilab 
(2002) Fermilab Mass Storage System -- Enstore. 
http://www.fnal.gov/docs/products/\hfill \break
enstore/html/intro.html [2003];\hfill \break
D.~Petravick,
in \textit{Proc. Int. Conf. Computing in High Energy and Nuclear Physics 
(CHEP 2000)}, \hfill \break
Padova, Italy, (7-11 Feb 2000) 630-633.



\end{thebibliography}
\end{document}